\documentclass[12pt,preprint]{aastex}

\newcommand\beq{\begin{equation}}
\newcommand\eeq{\end{equation}}
\newcommand\beqar{\begin{eqnarray}}
\newcommand\eeqar{\end{eqnarray}}

\newcommand\etal{{\it et al.~}}

\newcommand\OmegaM{\Omega_{\rm m}}
\newcommand\ngdot{\dot{n}_{\gamma,{\rm com}}}
 
\newcommand\iun{\mbox{${\rm \, photons \,\, cm^{-2} \, s^{-1}\,sr^{-1}}$}}
 
\begin{document}

\title{THE GUARANTEED GAMMA-RAY BACKGROUND}

\author{Vasiliki Pavlidou and Brian D. Fields}

\affil{Center for Theoretical Astrophysics, 
Department of Astronomy \\
University of Illinois, Urbana, IL 61801}

\begin{abstract}
The diffuse extragalactic gamma-ray background (EGRB) above 100 MeV
encodes unique information about high-energy processes in the
universe.  Numerous sources for the EGRB have been proposed, but the two
systems which are certain to make some contribution are active galaxies
(blazars) as well as normal galaxies.  In this paper, we evaluate the
contribution to the background from both sources. The active galaxy 
contribution arises from unresolved blazars. We compute this contribution
using the Stecker-Salamon model. For normal galaxies, the emission is 
due to cosmic-ray
interactions with diffuse gas.  Our 
key assumption here is that the cosmic-ray flux in a galaxy is
proportional to the supernova rate and thus the massive star formation
rate, quantified observationally by the cosmic star formation rate (CSFR).
In addition, 
the existence of stars today requires a 
considerably higher ISM mass in the past. Using the CSFR to compute both
these effects, we find that normal galaxies are responsible for
a significant portion ($\sim 1/3$) of the EGRB
near 1 GeV, but make a smaller contribution at other energies.
Finally, we present a ``minimal''
two-component model which includes contributions from both normal
galaxies and blazars.  We show that the spectrum of the diffuse
radiation is a key constraint on this model:
while neither the blazar spectra, nor the galactic spectra,
are separately optimal fits to the observed spectrum, the combined
emission provides an excellent fit.  
We close by noting key observational tests of this two-component model,
which can be probed by future gamma-ray observatories such as GLAST.
\end{abstract}


\section{Introduction}

All-sky $\gamma$-ray observations by SAS 2 \citep{fichtel,fichtel2} 
and most recently 
by EGRET \citep{sreek_obs}
have revealed the existence of an isotropic 
diffuse $\gamma-$ray emission, presumably
of extragalactic origin. This extragalactic gamma-ray background 
(EGRB) is well-described by a power-law energy spectrum with an index of 
$-2.10 \pm 0.03$, while the extragalactic intensity for energies 
$> 100 \ {\rm MeV}$ has an all-sky average value
$(1.45 \pm 0.05) \times 10^{-5} \iun $ 
\citep{sreek_obs,fichtel96}.

A variety of possible contributions to the EGRB
have been proposed. There are, however, two 
classes of $\gamma-$ray sources whose existence has been 
observationally established and thus guarantees that 
these make {\em some} contribution to the EGRB: blazars and normal galaxies. 

Blazars, which are active galactic nuclei with jets almost aligned with 
the line of sight
\citep{bk79,mgc92,dg95},
comprise the vast majority of the  identified  $\gamma-$ray point 
sources detected by EGRET 
\citep{egret3}. 
In addition, their energy spectra are power laws, with a 
distribution of indices peaked close to the 
EGRB energy spectrum index. It is therefore only logical to 
argue that a population of unresolved blazars 
with photon fluxes below EGRET sensitivity has to be  
the origin of a significant portion of the EGRB
\citep{ssm}.
A large AGN contribution 
to the EGRB has been anticipated as early as the 1970's 
(e.g., \citet{sww,bfht,pgfc}).
Given the EGRET results on blazars, \citet{ss94}
and 
\citet[henceforward SS96]{ss96}
made a detailed calculation of the blazar contribution to the
EGRB 
and indeed found it to be dominant, 
although the shape of the 
predicted blazar emission energy spectrum 
does not match the flatter 
spectrum of the latest EGRET EGRB determination \citep{sreek_obs}.

While the EGRET catalog of point sources is dominated by
blazars, the EGRET diffuse flux is dominated by   
emission from the Galactic plane. The latter is, 
for the most part, the result of the 
decay of neutral pions produced when cosmic rays interact 
with the interstellar medium.
The superposition of this diffuse $\gamma-$ray emission from all 
unresolved normal galaxies is the second guaranteed source of 
extragalactic background $\gamma-$ray intensity. The contribution 
of normal galaxies to the EGRB was calculated by
\citet{sww} for the case of non-evolving galaxies
and was found to be 
a few percent of the observed background.
\citet{lbp} extended this calculation to include
galactic evolution, which they found to 
significantly enhance the predicted background.
Their results spanned a range
$\Phi(> 100 \ {\rm MeV}) = 0.3 - 7 \times 10^{-6} \ \iun$
which comes much closer to the observed level.
\citet{schramm} and \citet{prantzos} inferred a
similarly large result from spallogenic element abundances.

{\it We define the sum of the $\gamma-$ray emission from 
all unresolved blazars and from all unresolved normal galaxies to 
be the guaranteed EGRB.} If the 
intensity level of the guaranteed EGRB can be confidently estimated, then 
by comparison to the observed EGRB one can constrain the 
observationally allowed contributions from any other hypothesized sources.

In this paper we present a new calculation of the contribution  
of normal galaxies to the EGRB.
We use observational estimates of the
cosmic star formation rate (CSFR), which have recently become 
available, to model the evolution of normal galaxy $\gamma-$ray emission.
To the latter, we then add the blazar component of the spectrum as 
given by SS96. Our results are computed for the 
currently favored $\Omega_\Lambda=0.7$, 
$\OmegaM=0.3$ cosmology. 
The resulting minimal two-component model will 
prove to be an excellent fit to the observed EGRET EGRB spectrum for 
energies up to 15 GeV, where $\gamma-$ray extinction is not important.

\section{Formalism}

The observable quantity which describes the EGRB is the differential intensity
$dI_E/d\Omega = dN_\gamma/dtdAdEd\Omega$. We need to calculate 
theoretically the differential intensity resulting from a collection of
unresolved $\gamma-$ray sources.
The differential intensity 
detected at the present cosmic epoch due to a population of 
$\gamma-$ray sources with collective comoving differential 
$\gamma-$ray emissivity density $\ngdot(z,E)
= dN_\gamma/dtdV_{\rm com}dE$ (where $V_{\rm com}$ is the comoving volume)
will be 
\beq \label{intro}
\frac{dI_E}{d\Omega} = 
\int \frac{\ngdot[z,(1+z)E]}{4 \pi a(t_0)^2r^2}
\frac{dV_{\rm com}}{dzd\Omega}dz \,\,,
\eeq
where $r$ is the coordinate 
distance of a source at a redshift $z$ and $a(t)$ is the dimensionless scale  
factor of the Friedmann - Robertson - Walker metric at a cosmic time $t$, 
so that $a(t_0)=1$. 
In a flat universe with 
matter density parameter $\OmegaM$ and vacuum energy density parameter 
$\Omega_\Lambda = 1 - \OmegaM$, eq.~(\ref{intro}) becomes
\beq \label{general}
\frac{dI_E}{d\Omega} = \frac{c}{4\pi H_0} 
\int 
\frac{\ngdot[z,(1+z)E]}{\sqrt{\Omega_\Lambda+\OmegaM (1+z)^3}}
dz \,\, ,
\eeq
where $H_0$ is the present value of the Hubble parameter.

In the case of blazars, SS96 have calculated $\ngdot$
considering blazars to be in either one of two states, flaring and
quiescent. Their calculation was done for an 
$\OmegaM=1$ cosmology. We have  confirmed that 
our blazar calculation exactly reproduces their result for
their chosen cosmology. We have then adapted 
their model to our preferred cosmology, keeping
all other parameters the same. We have confirmed that 
in the new cosmology, the derived blazar luminosity
function reproduces the EGRET 2nd catalog
flux distribution of observed blazars, in exactly the same
way as in the original Stecker-Salamon calculation. 

In the case of normal galaxies, 
$\ngdot$ can be expressed in terms of 
the CSFR function (mass being converted to stars per unit time per
unit comoving volume, denoted by $\dot{\rho}_\star(z)$).
The star formation 
rate (SFR, mass converted to stars per unit time) of an individual galaxy 
will be denoted by $\psi$. In order to associate the CSFR with 
$\ngdot$, we assume that:
(1) the high mass end of the initial mass function (IMF)
is universal, and thus $\psi$ as deduced from observations of high-mass
stars is always proportional to the 
supernova explosion rate in the same galaxy; 
(2) the cosmic ray flux in a galaxy is proportional to $\psi$ and
the cosmic ray spectral shape is universal (see \citet{focv});
and (3)
at any cosmic epoch 
the cosmic ray escape properties are the same as in the present Milky Way,
and any $\gamma$-rays produced after escape are negligible.

An average galaxy's $\gamma$-ray (number) luminosity is, by virtue of our 
assumptions, 
\beq
L_{\gamma}(z,E) = \frac{\psi(z)}{\psi_{\rm MW}} 
  \frac{\mu(z)}{\mu(0)}
  L_{\gamma, {\rm MW}}(E)
\eeq
where $E$ is the photon energy in the galaxy's rest frame,
and $\mu(z)$ is the gas mass fraction of the galaxy at redshift z. 
The factor 
$\mu(z)/\mu(0)$ has been introduced to account for the increase
of target atoms at earlier cosmic epochs and assumes 
a ``closed box'' galaxy.
The emissivity density will then be 
\beq\label{ng}
\ngdot(z,E)=
L_\gamma n_{\rm gal} = 
L_{\gamma, {\rm MW}}(E) \frac{\dot{\rho}_\star(z)}{\psi_{\rm MW}}
\frac{\mu(z)}{\mu(0)} 
\,\, ,
\eeq
where the comoving galaxy number density
$n_{\rm gal}$ and star formation rate $\psi(z)$ are
related to the CSFR via $\psi n_{\rm gal} = \dot{\rho}_\star$.
Note that, due to our assumptions, the conversion 
of a certain amount of gas into stars will result to the production
of the same amount of $\gamma$ rays from CR-ISM interactions 
{\em regardless of the way the star formation is distributed 
among galaxies}. Therefore, our calculation does not depend 
on the observational knowledge of both $\psi$ and $n_{\rm gal}$
individually, but only on that of their product, which is 
the CSFR.

Now the sum of gas mass and star 
mass in a closed box is constant in time, 
equal to the baryonic mass of the galaxy.
Hence, 
\beq \label{mu}
\mu(z) = 1 - \left[ 1-\mu(0) \right] 
	\frac{\int_{z_\star}^z  \ dz
\frac{dt}{dz} \dot{\rho}_\star(z)}
{\int_{z_\star}^{0} \ dz \frac{dt}{dz} \dot{\rho}_\star(z)}
\,\, ,
\eeq
where $z_\star$ is the assumed redshift at which star formation begins.

Equations (\ref{general}), (\ref{ng}), and (\ref{mu}) 
can be combined to give
\begin{eqnarray} \label{dido}
\frac{dI_E}{d\Omega} & = & \frac{c}{4\pi H_0 \psi_{_{\rm MW}}} 
\int_0 ^{z_\star} \! \! \! \! dz \left\{
\dot{\rho}_\star(z)
\frac{L_{\gamma, _{\rm MW}}[(1+z)E]}
{\sqrt{\Omega _\Lambda+\OmegaM (1+z)^3}}
\right. 
\nonumber \\ 
& &  \left. \left[ \frac{1}{\mu_{\rm 0,_{\rm MW}}}  
\mbox{} - \left( \frac{1}{\mu_{\rm 0,_{\rm MW}}}-1 \right) 
	\frac{\int_{z_\star}^z \ dz \frac{dt}{dz} \dot{\rho}_\star(z)}
             {\int_{z_\star}^0 \ dz \frac{dt}{dz} \dot{\rho}_\star(z)}
\right] \right\}
\,\, . 
\end{eqnarray}
Here $\psi_{\rm MW}$ is the star formation rate of the MW today and 
$\mu_{\rm 0,MW}$ is the MW gas mass fraction today. 

\section{Inputs}

\subsection{Cosmic Star Formation Rate}

The observational determination of the CSFR is based on 
observations of the cosmic luminosity density in 
various wavebands which are dominated by emission 
from short-lived stars; these include the UV 
(e.g., \cite{lilly96, madau98}), 
the IR (e.g., \cite{chary01, cole01})
and H$\alpha$ 
(e.g., \cite{gallego95, tm98, hopkins00}). 
For $z<1$ there is a relatively
clear picture of the CSFR evolution, and results from different 
wavebands generally agree that the CSFR increases to peak around 
$z=1$ reaching a maximum  more than an 
order of magnitude larger than 
its present value. For $z = 1-4$, however, the uncertainty in the 
CSFR reaches about an order of magnitude, as 
the derived CSFR values depend strongly on the assumed 
amount of dust extinction
(see, e.g., the compilation of CSFR estimates using different tracers 
by \cite{hopkins01}). 

In computing the normal galaxy contribution to the EGRB we will use
the analytic fit of the CSFR evolution given by \cite{cole01}, 
based on 
near-infrared observations, a Salpeter mass function and
an $\Omega_\Lambda=0.7$, $\OmegaM =0.3$ cosmology.
We will refer to their fit of data points not corrected
for dust extinction as the ``uncorrected CSFR'' and to their fit 
of data points corrected for dust absorption 
as the ``dust-corrected CSFR.''

\subsection{The Observed Galactic Gamma-Ray Spectrum}

In performing the intergal of eq.~(\ref{dido}) we will also 
need the functional form of the differential diffuse
$\gamma-$ray (number) luminosity of the Milky Way, $L_{\gamma, {\rm MW}}$.
We will use EGRET observations of the $\gamma-$ray 
{\em flux}
from the Galactic plane to deduce the shape of the spectrum.
We find that the EGRET flux spectrum can be well fitted by a double power
law, of spectral indices $-1.52$ for energies below $850 {\rm \, MeV}$
and $-2.39$ for higher energies.
The normalization of the (number) luminosity 
spectrum can be determined from the requirement that 
$
\int^\infty _{\rm 100 \, MeV} L_{\rm \gamma, {\rm MW}}(E) dE
= q_{\gamma}(>100 {\rm \, MeV}){\cal N}_{\rm H, MW}
$
where $q_{\gamma}(>100 {\rm \, MeV})$ is the total $\gamma-$ray 
emissivity per hydrogen atom and ${\cal N}_{\rm H, MW}$ is the number of
H atoms in the MW. Using a value of 
$q_{\gamma}=2.4 \times 10^{-25} 
  \ {\rm photons \ s^{-1} \ (H \ atom)^{-1}}$ (see, e.g., \citet{pf1} and
references therein)
and a total gas content of the MW $M_{\rm gas, MW} \sim 10^{10}
 {\, \rm M_\odot}$ (see, e.g., Fields \etal 2001 for a discussion of
the uncertainties involved in this estimate) we find $
\int^\infty _{\rm 100 \, MeV} L_{\rm \gamma, {\rm MW}}(E) dE = 
2.85 \times 10^{42}$ photons ${\rm s^{-1}}$. 
This gives
$L_{\gamma, _{\rm MW}}(E) =  7.21 \times 10^{38} (E/{\rm 850 \, MeV})
^{-\kappa}$ $\gamma \ {\rm s^{-1} \ MeV^{-1} }$, where
$\kappa=1.52$
for  $E \le 850$ MeV
and
$\kappa=2.39$
for  $E > 850$ MeV.

\subsection{Other Inputs}

The MW star formation rate 
is a poorly known quantity, and estimates for its value
range from $1.6 {\rm M_\odot \, yr^{-1}}$ \citep{tan00} to 
$3.5 {\rm M_\odot \, yr^{-1}}$ (derived from the estimate of 
\cite{drag99} for the MW supernova rate 
using a Salpeter IMF and a supernova progenitor mass cutoff 
of $8 \,{\rm M_\odot}$). Here, we will use 
$\psi_{\rm MW} = 3.2 {\rm \, M_\odot \, yr^{-1}}$ 
\citep{mckee89}, which lies in the upper range of the available estimates
and therefore leads to a conservative estimate of the normal galaxy EGRB
component. 
For the MW gas mass fraction today we will adopt 
$\mu_{\rm 0,MW}=0.14$, using $M_{\rm gas, MW} = 10^{10} {\rm
M_\odot}$ and a
star mass value of $M_{\rm \star, MW} = 7 \times 10^{10}$
\citep{pagel}.
Finally, we will use $z_\star=5$ for the redshift for which star formation
begins.

\section{Results}

\begin{figure}
\plotone{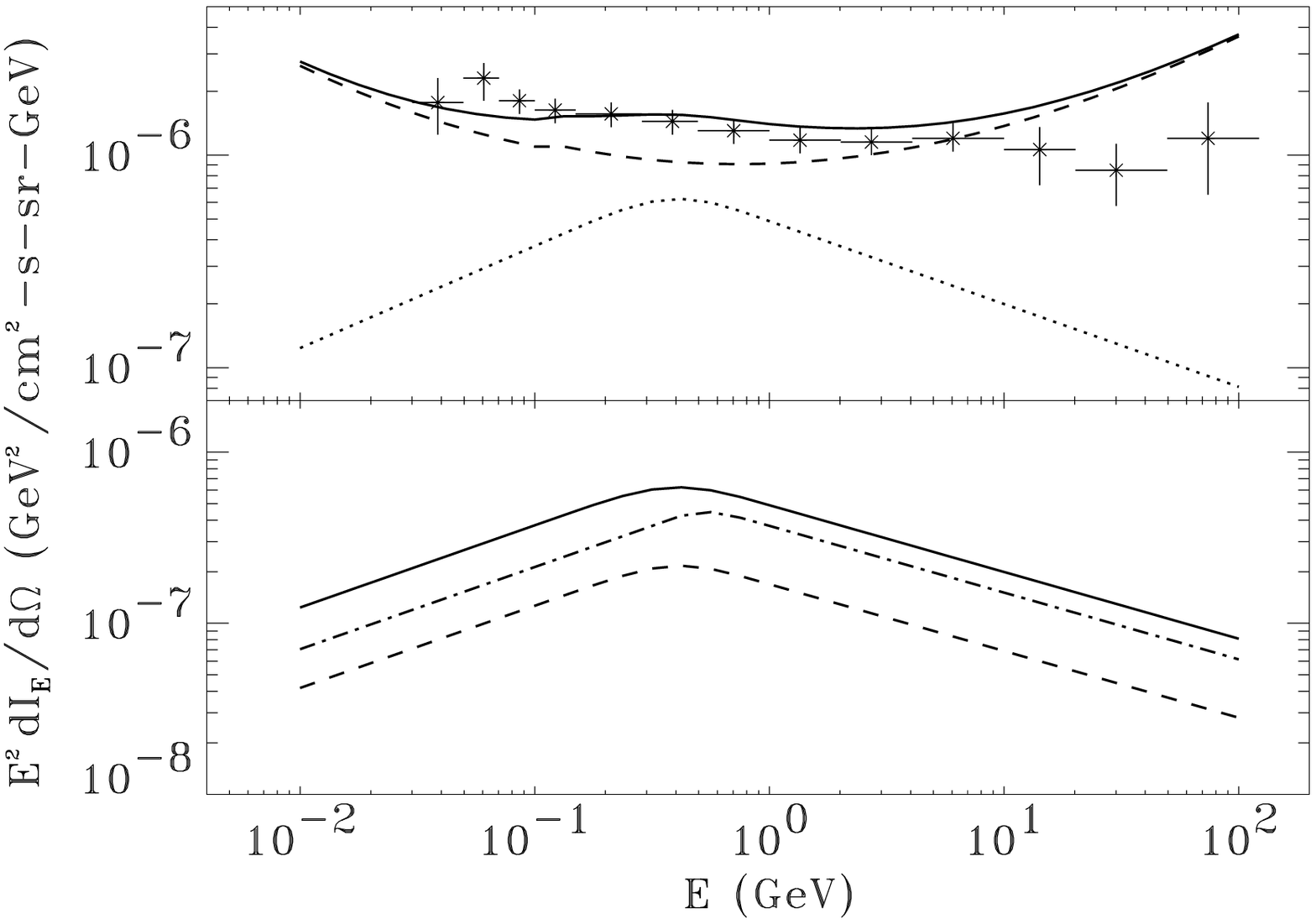}
\caption{
Upper panel: Blazar (dashed line) and normal galaxy
(dotted line) contributions to the EGRB, overplotted with 
the
summed ``minimal model'' spectrum (solid line)
and the EGRET data points from \cite{sreek_obs}. Lower 
panel:
Normal galaxy spectrum for a dust-corrected CSFR (solid 
line),
uncorrected CSFR (dashed line) and dust-corrected CSFR but 
with the
integration only extending up to $z_\star=1$ (dot-dashed 
line).}\label{fig1}
\end{figure}

The spectrum of the normal galaxy contribution to the 
EGRB, as derived from eq.~(\ref{dido}), and for a dust-corrected
CSFR, is plotted in the upper panel of Fig. \ref{fig1}. 
In the same plot, we have overplotted the blazar 
contribution as calculated from the 
SS96 model, for our
preferred $\Omega_{\Lambda}=0.7$, $\OmegaM=0.3$
cosmology, as well as the 
``minimal'' two-component model of the 
guaranteed EGRB. This summed spectrum 
has a flatter shape than either of its constituent
spectra due to the fact that the maximum of the 
convex normal galaxy curve happens to lie in 
the same energy regime with the minimum of the
concave blazar spectrum. 

The ``minimal model'' 
is in excellent agreement   
with the observational data points 
from EGRET \citep{sreek_obs}, both in 
amplitude and in spectral shape, for energies up to 
15 GeV. For higher energies, absorption effects 
due to pair production, which have not been 
treated here, become important\footnote{
The threshold for 
intergalactic absorption or downscattering by
pair production is at
$E_{\gamma,{\rm th}} = m_e^2/\epsilon_{\gamma,{\rm bgnd}} 
  = 0.21 \ {\rm TeV} \ (\lambda_{\gamma,{\rm bgnd}}/1 \ {\rm \mu m})$.
The photon backgrounds with enough energy density to
introduce a significant cutoff are the UV, optical, and IR
\citep{mph96,ss98}, so that the practical cutoff is at about
$\sim 0.075 \ {\rm \mu m}$ (Madau \& Pozzetti 2000, taking their
$\sim 0.15 \ {\rm \mu m}$ observed cutoff to correspond to $z \sim 1$).
Therefore, the threshold is  
$E_{\gamma,{\rm th}} \sim 15$ GeV, which is interestingly 
the energy of the first data point deviating from the 
computed spectrum. For energies above that 
threshold, the reader should be aware that our spectra should 
be reduced by about a factor $\sim 2$;
we plan to return to this effect in a future publication.}.
For the first 10 points with energies $\la 15$ GeV, the minimal 
model has a $\chi^2$ per degree of freedom $\nu$ of
$\chi^2_\nu = 0.78$, indicating an excellent fit. For comparison, the 
blazar contribution alone has $\chi^2_\nu =4.68$.
Had we made use of the free parameter in the  
Stecker-Salamon blazar model which allows for the variation of  
the total amplitude of the blazar contribution, 
the blazar-only model has a minimum $\chi^2_\nu = 0.95$, 
still somewhat larger than that of the summed 
normal galaxy - blazar spectrum. Moreover, such an increase 
of the blazar contribution would be ignoring the guaranteed
contribution from normal galaxies which, if added to the 
boosted blazar signal, would overpredict the EGRB. Therefore, 
we conclude that no modification is needed in the Stecker \& Salamon 
(1996) calculation for energies $\la 15$ GeV.

Uncertainties in our normal galaxy spectrum calculation
are introduced due to uncertainties in the determination 
of $\dot{\rho}_\star(z)$, $M_{\rm gas, MW}$, $\mu_{\rm 0,MW}$
and $q_{\gamma,_{\rm MW}} $.
Of our input parameters, $\psi_{\rm MW}$ and $q_{\gamma,_{\rm MW}} $
enter our calculation as multiplicative factors and therefore
uncertainties in their values do not affect the shape of the spectrum, 
but only the overall normalization (introducing an overall uncertainty
of a factor $\sim$ 4). Our results are relatively insensitive to the value
of $M_{\rm gas, MW}$ since a change in its value affects the 
calculation through 
$L_{\gamma, {\rm MW}}$ and $\mu_{0, {\rm MW}}$ in opposite directions. 
In addition, our calculation shows that most of the
background intensity in the normal galaxy component originates from $z<1$.
Therefore, our results are not affected significantly by the CSFR
evolution at $z>1$ where the CSFR uncertainties can reach an order of
magnitude. This fact is demonstrated
in the lower panel of Fig. \ref{fig1}, where 
we have plotted the normal galaxy
spectrum for both the dust-corrected CSFR
and the uncorrected CSFR, together with 
the spectrum derived for a dust-corrected CSFR 
in the extreme case where  {\em no} star formation 
is assumed to have
taken place at $z>1$. 
The difference from the full 
integration up to $z=5$ is less than a factor of 2.
We note that in the latter case, the peak of the spectrum is displaced
towards higher energies. On the other hand, if the CSFR was much higher at
high redshifts, as suggested recently by \citet{lanzetta}, this would
displace the peak of the spectrum towards lower energies. 

\section{Discussion}

The minimal 2-component model of the EGRB can be tested and 
improved in various ways when observations from future $\gamma$ -
ray telescopes such as GLAST 
become available.
On the one hand, the improved effective collecting area of 
GLAST as compared with that of EGRET, especially at high energies, 
will allow GLAST to accurately measure the shape of the EGRB for 
energies up to 1 TeV. This will test whether the spectrum turns over at 
energies higher than $\sim 15$ GeV due to electron-pair production
via interactions with the IR, UV and optical backgrounds
\citep{mph96,ss98}. 
Any 
model in which the EGRB is assumed to be of cosmic origin 
should exhibit this behavior. If the GLAST data do not 
confirm that prediction, all extragalactic models are ruled out 
unless they can predict compensatingly steepening spectra with 
increasing redshift. 

On the other hand, the improved point source sensitivity of 
GLAST will allow it to resolve a higher number of blazars 
($\sim $ 100 more than EGRET, Stecker \& Salamon 1999), and 
therefore the blazar contribution to the EGRB will be reduced by 
about a factor of 2. 
If unresolved blazars are the only constituent of the EGRB, the 
{\it fractional change} of the EGRB  will be the same as the 
fractional change of the background blazar emission. If, however, 
there is a second component in the EGRB (in our case, that of 
normal galaxies), the fractional change of the EGRB 
should be smaller.

In addition, with the blazar component reduced by a factor of 2, our
calculated normal galaxy contribution will become comparable to that of
blazars for energies $\sim 1$GeV. Therefore, if the relative contributions
of blazars and normal galaxies to the minimal model are comparable to our
estimates, the shape of the EGRB spectrum should start to exhibit a
(detectable in principle) deviation from its single power-law form at
$\sim 1$ GeV, corresponding to the normal galaxy spectrum peak. 
Were this peak detected, the relative contribution of normal galaxies to
the EGRB could be determined observationally.
 
The observations of GLAST can also be used to improve 
the minimal model and its predictions. The observations 
of more blazars will allow a more confident determination 
of the observational inputs for the SS96 blazar model, 
as pointed out by \citet{ss99}.
As far as the normal galaxy  component model is concerned, 
GLAST is expected to detect several Local Group galaxies
(the SMC, LMC, M31 and maybe M33; Pavlidou \& Fields 2001), 
and therefore it will be possible to check our assumption 
of the universality of the galactic diffuse gamma-ray 
emission spectrum.

Finally, with both guaranteed EGRB components well-understood, 
one can better identify or constrain any other 
components and any new physics which might generate them.

\acknowledgements{We thank Kostas Tassis for 
enlightening discussions, and the referee, O.C. de Jager,
for helpful suggestions which improved this paper. This work was supported by 
National Science Foundation grant AST-0092939.
The work of VP was partially supported 
by a scholarship from the Greek State Scholarships Foundation.}

\clearpage

\end{document}